\documentclass[aps,prl,twocolumn,showpacs,superscriptaddress]{revtex4}

\usepackage{amssymb}
\usepackage{graphicx}
\usepackage{xcolor}
\usepackage{amsmath}
\usepackage{amsfonts}
\usepackage{amssymb}
\usepackage{braket}

\begin{document}

\title{Image-charge detection of the Rydberg states of surface electrons on liquid helium}

\author{Erika Kawakami}
\email[E-mail: ]{erika.kawakami@oist.jp}
\affiliation{Quantum Dynamics Unit, Okinawa Institute of Science and Technology, Tancha 1919-1, Okinawa 904-0495, Japan}
\affiliation{PRESTO, Japan Science and Technology (JST), Kawaguchi, Saitama 332-0012, Japan}
\author{Asem Elarabi}
\affiliation{Quantum Dynamics Unit, Okinawa Institute of Science and Technology, Tancha 1919-1, Okinawa 904-0495, Japan}
\author{Denis Konstantinov}
\email[E-mail: ]{denis@oist.jp}
\affiliation{Quantum Dynamics Unit, Okinawa Institute of Science and Technology, Tancha 1919-1, Okinawa 904-0495, Japan}
\date{\today}

\begin{abstract}
We propose and experimentally demonstrate a new spectroscopic method, image-charge detection, for the Rydberg states of surface electrons on liquid helium. The excitation of the Rydberg states of the electrons induces an image current in the circuit to which the electrons are capacitively coupled. In contrast to the conventional microwave absorption measurement, this method makes it possible to resolve the transitions to high-lying Rydberg states of the surface electrons. We also show that this method can potentially be used to detect quantum states of a single electron, which paves a way to utilize the quantum states of the surface electrons on liquid helium for quantum computing.                 
\end{abstract}


\maketitle

Surface electrons (SE) above liquid helium constitute an exquisitely pure quantum system which for a long time served as a unique experimental platform to discover interesting many-electron phenomena~\cite{Andrei-book}. The quantized (Rydberg) states of SE with a hydrogen-like energy spectrum $E_n=-R_e/n^2$, where $R_e\sim 10^{-3}$~eV and $n$ is a positive integer number, are formed due to the attractive interaction between an electron and its image charge inside the liquid, as well as a strong repulsive barrier experienced by the electron at the vapor-liquid interface. For typical experimental temperatures below 1 K, electrons occupy the ground state and are localized about 10~nm above the surface. The higher-energy Rydberg states can be excited by millimeter-waves ($\sim 100$~GHz) radiation.  Grimes and Brown first measured the transitions from the ground ($n=1$) to the low-lying excited states ($n=2\sim 6$) by detecting the change in the microwave (MW) absorption caused by the excitation of the Rydberg states of SE using a cryogenic bolometer~\cite{GrimesPRL1974}. A renewed interest in the Rydberg states of SE has emerged from their potential as qubit states~\cite{PlatzDykm1999,DykmPRB2003}, which was followed by several other proposals to use either orbital or spin states of SE as qubit states~\cite{LyonPRA2006,SchuPRL2010}. A crucial point for successful qubit implementation is the ability to manipulate and detect quantum states of a single electron. For such an application, the conventional MW absorption measurement is inappropriate as a detection of the quantum state because it is applicable only for a sufficiently large number of electrons~\cite{GrimesPRL1974,LambPRL1980,CollPRL2002}, and so is the indirect detection of the Rydberg transitions via the measurement of SE conductivity~\cite{VolJETP1981,KonsPRL2007}. Originally, a destructive readout of the Rydberg states was proposed, in which the electrons leave the liquid surface depending on the occupied Rydberg states~\cite{PlatzDykm1999,WillPRL1971}. An interesting and promising idea is to use the strong coupling of a single electron to a superconducting resonator to realize a non-destructive readout of electron quantum states~\cite{SchuPRL2010,GePRX2016}. However, this method is limited to a low transition frequency which should match the frequency of the coplanar resonator ($\sim 5$~GHz), thus it is not applicable for the detection of the excitation of the Rydberg states.

Here, we propose and demonstrate a new spectroscopic method, image-charge detection, for the Rydberg states of SE on liquid helium. Our method makes use of the fact that, as an electron is excited from the ground state to a higher excited state, the electron wave function spreads farther away from the liquid surface and its average distance from the liquid surface increases. Thus, when the system is placed near an electrode aligned parallel to the liquid surface, the excitation of SE causes a change in the image charge induced in the electrode by SE. In our experiment, the method is demonstrated with MW-excited SE placed in a parallel-plate capacitor. By measuring the image current in the capacitor, we managed to detect the excitation to the high-lying Rydberg states which were unable to be measured with the above-mentioned conventional methods. We also discuss an alternative detection scheme where the electric susceptibility of the MW-excited SE is detected as a relative change in the capacitance, a technique which can be scaled down to detection of the excitation of a single electron.
\begin{figure}[tbp]
\begin{center}
\includegraphics[width=8.5cm]{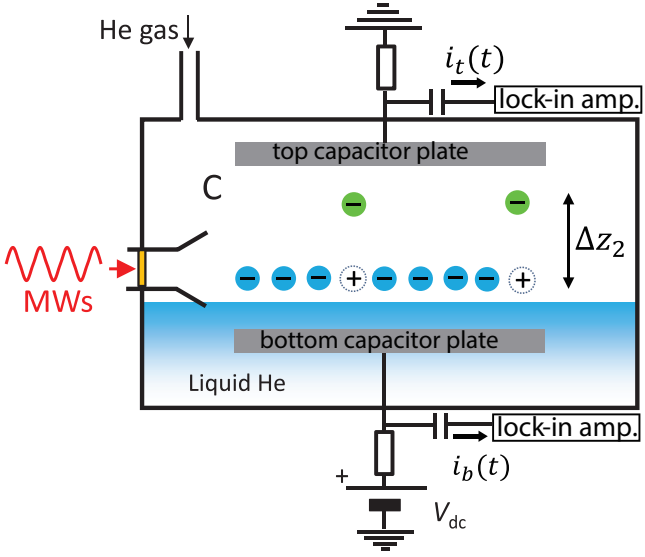}
\end{center}
\caption{(color online) Principle diagram of the excited-state-population detection of the electrons trapped on the surface of liquid helium (light blue circles) placed between top plates and bottom plates of a capacitor C. The electrons that are excited to the first excited state (green circles) are elevated above a charged layer of the ground-state SE by a distance $\Delta z_{2} \sim 35$~nm and forms an electric dipole $p=e\Delta z_{2} $. In the experiments shown here, an induced image current $i(t)$ was detected using a lock-in current amplifier by periodically varying $i$ by means of the modulated MW excitation. A suitable bias-tee is placed at room temperature for each  capacitor plate (see text for more details).}
\label{fig:principle_diagram}
\end{figure}

The principle diagram of the image-charge detection is shown in Figure~\ref{fig:principle_diagram}.
SE are formed on the surface of liquid helium which are placed between two plates of a capacitor C. Here we use a parallel-plate capacitor (the area of each plate $S$ and the distance between the plates $D$). SE are confined on the surface of liquid helium by applying a positive dc bias $V_\textrm{dc}$ to the bottom capacitor plate. Owing to the linear Stark shift of the Rydberg levels caused by a dc electric field $E_\perp$ applied perpendicular to the surface, the transition frequency of SE, $\omega_{1n}(E_\perp)=(E_n - E_1)/\hbar$, can be adjusted to match with the microwave frequency $\omega_0$ by varying the value of $V_\textrm{dc}$. The MW-excited electrons are elevated above the charged layer of the ground-state electrons by a distance $\Delta z_{n} = z_{nn} - z_{11}$, where $z_{nn}$ is the average value of $z$-coordinate (counted from the surface) of an electron occupying the Rydberg state of index $n$. For the first excited state, $n=2$, this distance is $\Delta z_{2}  \approx 35$~nm, and it increases with $n$. As a fraction $\rho_{nn}$ of SE is excited to the $n$-th Rydberg state, the image charge induced by SE in the top (bottom) capacitance plate changes by $\Delta q$ ($-
\Delta q$):

\begin{equation}
\Delta q =  \frac{\Delta z_{n} }{D} e n_s \rho_{nn} S   =\frac{\Delta z_{n} }{D} P_e S ,
\label{eq:deltaq}
\end{equation}
\noindent where $e(>0)$ is the elementary charge and $n_s$ is the areal density of SE. Here, for later discussion, we introduce the quantity $P_e =n_s S \rho_{nn} p/ (S \Delta z_{n} )=en_s  \rho_{nn}$, with $p=e \Delta z_n$ being the electric dipole moment of one electron. This quantity can be viewed as the electric dipole moment per unit volume of the electron system induced by the excitation of SE to the $n$-th excited state. The change in the image charge induces a current $i_{t,b}$ in the top (bottom) capacitor plate: 

\begin{equation}
i_{t,b}(t)=\pm \frac{d\Delta q}{dt}.
\label{eq:i}
\end{equation}

\noindent This current can be readily detected using, for example, a lock-in amplifier by periodically varying the fractional occupancy $\rho_{nn}$.  The periodic variation of $\rho_{nn}$ can be realized by MW modulation or by the modulation of the voltage applied to the capacitor plates. In the latter case, $\rho_{nn}$ is determined by the detuning which is related to the electric field $E_\perp$ applied to SE by $\delta \omega = \omega_{1n}(E_\perp) - \omega_0$. This case is discussed later as a future experiment and the former is realized by applying a pulse-modulated MW excitation to the system in the experiment described here. Assuming the harmonic time-dependence of $\rho_{nn}$ at the pulse modulation frequency $\omega_m$: $\rho_{nn}=\rho_{nn}^{(0)} e^{i \omega_m t}$, the current amplitude can be estimated as

\begin{equation}
|i_{t,b}(t)|  = \left| \frac{d\Delta q}{dt} \right| = \frac{ e n_s C_0\omega_m  \Delta z_{n} \rho_{nn}^{(0)}}{ \varepsilon_0},
\label{eq:i_PM}
\end{equation}
 
\noindent where $C_0=\varepsilon_0 S/D$ is the the capacitance of the parallel plate capacitor and  $\varepsilon_0$ is the vacuum permittivity. 
For typical values of $n_s=10^8$~cm$^{-2}$, $C_0\sim 1$~pF, the modulation frequency $\omega_m/2\pi=100$~kHz, and $\rho_{nn}^{(0)}=10\%$, we obtain $|i_{t,b}(t)|\sim 100$~pA. 

The experiment is carried out in a leak-tight copper cell cooled below 1~K in a dilution refrigerator and filled with condensed He gas. In the experiments shown here, we used liquid $^3$He. Similar results were obtained using liquid $^4$He. The helium liquid surface is set approximately midway between two plates of a parallel-plate capacitor formed by two round (diameter 24~mm) conducting plates separated by four quartz spacers of height $2$~mm. This height sets the gap $D$ between the capacitor plates. Using a parallel-plate capacitor with the gap smaller than used here is not practical because the helium liquid, which has the capillary length $\sim 0.5$~mm, would wet the top plate by the capillary rise and form a meniscus~\cite{CollPRB2017}. Each plate of the capacitor consists of three concentric electrodes separated by two gaps of diameters 14 and 20~mm and width 0.2~mm. Electrons are produced by the thermionic emission from a tungsten filament placed near the capacitor and above the liquid surface, and electrons are attracted towards the liquid surface by applying a positive bias to all the three concentric electrodes of the bottom plate. After the SE system is formed on the surface above the bottom plate, the areal  density of SE is determined by the voltage bias from the condition of complete screening of the electric field above the surface. After the system is formed, the voltage of the outer concentric electrode (guard) is set to a large negative value (typically -30~V) to confine SE above the two inner concentric electrodes and ensure that electrons do not escape to the grounded walls of the cell. The two inner electrodes of each plate constitute a Corbino disk which allows detection of SE by driving one of the electrodes with a small ac voltage and measuring the current induced in the other plate by the lateral motion of SE~\cite{SommPRL1971}. MW radiation at a fixed frequency $\omega_0$ is introduced into the cell through a sealed rectangular single-mode waveguide with inner dimensions $1.6\times 0.8$~mm. In order to increase the area illuminated by MW, the waveguide gradually transforms to an overmoded size ($3.8\times 1.9$~mm) inside the cell (shown schematically in Fig.~\ref{fig:principle_diagram}). 

\begin{figure}[tbp]
\begin{center}
\includegraphics[width=7.5cm]{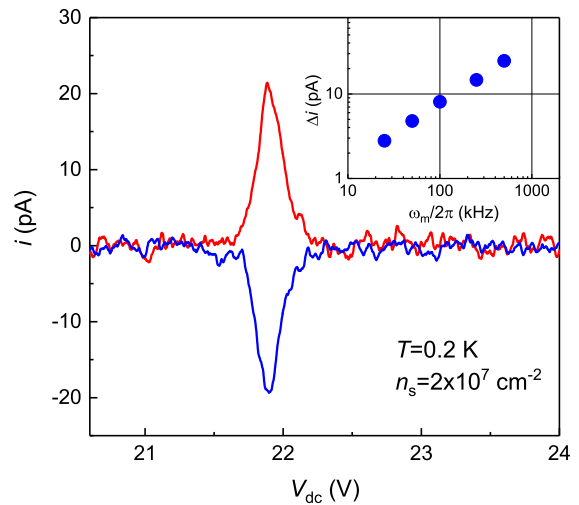}
\end{center}
\caption{(color online) The current signal measured at the bottom (red line) and top (blue line) capacitor plate for SE irradiated with pulse-modulated MW at frequency $\omega_0=140$~GHz. (Inset) The current signal at the resonance measured for different values of the pulse-modulation frequency $\omega_m$.}
\label{fig:2}
\end{figure}

\begin{figure}[tbp]
\begin{center}
\includegraphics[width=7.5cm]{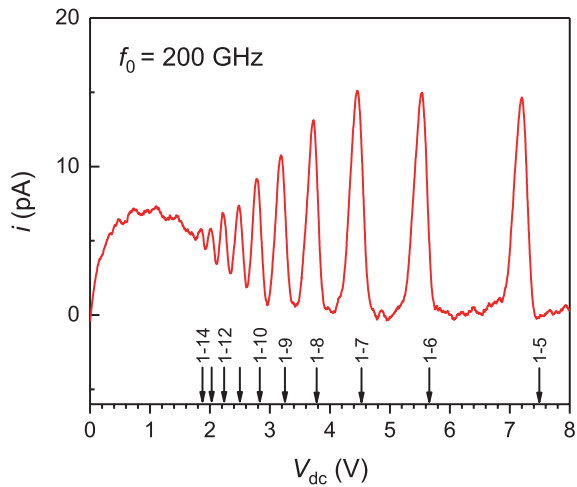}
\end{center}
\caption{(color online) The current signal measured at the bottom capacitor plate for SE irradiated with pulse-modulated MW at frequency $\omega_0/2\pi=200$~GHz while bias $V_\textrm{dc}$ applied to the bottom electrode is swept down to zero. The arrows indicate theoretical predictions for the transitions between the ground states and excited states. }
\label{fig:4}
\end{figure}

A resonant $1\rightarrow n$ transition between the Rydberg states of SE is excited by adjusting voltage $V_\textrm{dc}$ applied to the Corbino disk of the bottom plate to match $\omega_{1n}$ with $\omega_0$ via Stark shift. In the experiment, we employ on/off MW pulse modulation at frequency $\omega_m$ and measure the corresponding ac currents at the Corbino disk of either the bottom or top capacitor plates using a lock-in amplifier, while sweeping  $V_\textrm{dc}$ through the resonance. The typical result of such measurements is shown in Fig.~\ref{fig:2} where we plot the in-phase component of the current measured by the lock-in amplifier at the modulation frequency $\omega_m/2\pi=250$~kHz. The sharp enhancement of current corresponds to the resonant $1\rightarrow 2$ transition of SE excited by the applied MW at frequency $\omega_0=140$~GHz. The resonance value of $V_\textrm{dc}$ corresponds to the electric field acting on the electrons $E_\perp=V_\textrm{dc}/D\approx 110$~V/cm, which agrees reasonably well with the calculated value $116$~V/cm for the Stark-shifted Rydberg levels assuming an infinitely-large barrier for the electrons at the vapor-liquid interface. Possible causes for this deviation, as well as for the deviation seen in Fig.~\ref{fig:4} discussed later, come from the approximate model used to calculate the Rydberg spectrum and the contribution to $E_\perp$ from the image charges induced by SE in the capacitor plates. The full width of the peak at its half-maximum is about $400$~MHz, which is significantly larger than the expected intrinsic linewidth $\gamma\approx 1$~MHz due to the elastic scattering of SE from ripplons~\cite{IsshJLTP2007}. This width is determined by the inhomogeneous broadening of the transition energy due to the nonuniformity of $E_\perp$, which most likely arises from the misalignment between the capacitor plates and liquid surface. 

An important feature of the result shown in Fig.~\ref{fig:2} is that the current signals measured at the top and bottom plates are nearly equal in magnitude but opposite in sign as expected from Eq.~(\ref{eq:i}). The inset of Fig.~\ref{fig:2} shows a nearly linear dependence of the magnitude of $i$ at the resonance for different values of $\omega_m$, as expected from Eq.~(\ref{eq:i_PM}).

\begin{figure}[tbp]
\begin{center}
\includegraphics[width=7.5cm]{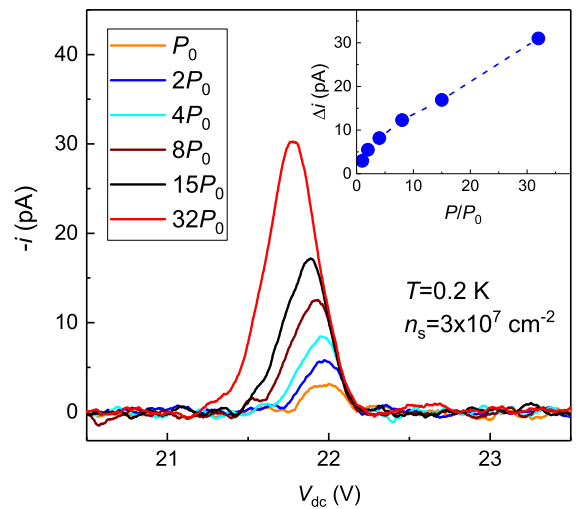}
\end{center}
\caption{(color online) The current signal measured at the top capacitor plate for SE irradiated with pulse-modulated MW at frequency $\omega_0/2\pi=140$~GHz for different MW power. Inset: The current signal at the resonance versus the normalized MW power.}
\label{fig:3}
\end{figure}   

Figure~\ref{fig:3} shows the dependence on MW power of the current signal measured at the top plate for the resonant $1 \rightarrow 2$ transition. The magnitude of the current at the resonance versus normalized power is shown in the inset of this figure. For a two-level system, the fractional occupancy $\rho_{22}$ is expected to increase linearly for small powers and saturate at the value $0.5$ at high powers. The measured dependence shows a  deviation from such a dependence. It is known that in SE the two-level approximation fails for a high MW power due to the electron heating by MW radiation, which results in the thermal population of the Rydberg states which are not directly excited by MW~\cite{KonsPRL2007}. Thus, the expression $\Delta z_{n} \rho_{nn}^{(0)}$ in Eq.~(\ref{eq:i_PM}) becomes reformulated as $\sum_{k \geq n} \Delta z_{k} \rho_{kk}^{(0)}$. The data in Fig.~\ref{fig:3} shows that the resonance shifts towards lower values of $V_\textrm{dc}$ with increasing power. This is due to the Coulomb shift of the Rydberg levels caused by the electron-electron interaction~\cite{KonsPRL2009}. Using the experimentally determined $\omega_{12}(V_\textrm{dc})\sim 2.1$~GHz/V, the estimated Coulomb shift is about 0.2~GHz for the highest applied MW power (red line in Fig.~\ref{fig:3}). Assuming the Boltzmann population for all the Rydberg states, the corresponding estimated electron temperature is $T_e\approx 5$~K~\cite{KonsPRL2007}. The thermal population of the excited states with higher quantum numbers should further increase the image current $i$, thus a stronger rise of $i$ with $P$ should be expected at higher $P$. 

The fact that the image current induced by the excitation of SE strongly increases with quantum number $n$ of the excited state suggests one of the important advantages of our method comparing with the conventional MW absorption measurement. The MW absorption due to the resonant $1\rightarrow n$ transition is proportional to the oscillator strength $f_{1n} \propto\hbar \omega_{1n} |\braket{1|z|n}|^2 $. The Thomas-Reiche-Kuhn sum rule, $\sum\limits_n f_{1n}=1$, states that the oscillator strength, and therefore the absorbed power, must decrease rapidly with increasing $n$ in order to ensure a convergence of the sum. Indeed, in a typical power absorption measurement, the measured signal decreases rapidly with increasing $n$ and for large $n$ the transitions become practically unobservable~\cite{GrimPRB1976}. In our method, the rapid decrease of the $1\rightarrow n$ transition rate $|\braket{1|z|n}|^2$  with increasing $n$ is compensated by an increase of $\Delta z_n$. Figure~\ref{fig:4} shows the current signal measured at a fixed MW power as  $V_\textrm{dc}$ is swept to zero, i.e., tuning transitions to higher-$n$ states to the resonance with the applied MW.  We can clearly observe the transitions up to $n=14$. At lower $V_\textrm{dc}$, since the Stark shift of the Rydberg levels is smaller, the overlap between different transitions becomes larger and produces a smooth background, which makes it difficult to resolve each transition. However, this is not a fundamental limitation. High-lying Rydberg states can be resolved by increasing MW frequency $\omega_0$, thus shifting resonances to higher  $V_\textrm{dc}$. 

In order to increase the sensitivity and the bandwidth of the image-charge  detection, a number of improvements can be readily done, such as an employment of a cryogenic high-electron-mobility transistor (HEMT) amplifier~\cite{VinkAPL2007,AschRSI2002}. Another possibility, as mentioned earlier, is to apply a small ac voltage $u(t)$ to a capacitor plate in addition to the dc bias $V_\textrm{dc}$ to induce the periodic variation of the fractional occupancy $\rho_{nn}$. In this case, we can draw an analogy between the capacitance change induced by inserting a dielectric slab into a capacitor and that by the MW-excited SE. We introduce the electric field $\mathcal{E}=u/D$ due to the ac voltage applied to the capacitor and, for the sake of simplicity, ignore the effect of the electric field from the image charges induced by SE in the capacitor plates. By making expansion of $\rho_{nn}(u)=\rho_{nn}|_{u=0}+\frac{d\rho_{nn}}{d \mathcal{E}}|_{u=0}\mathcal{E}+\frac{d^2\rho_{nn}}{d \mathcal{E}^2}|_{u=0}\frac{\mathcal{E}^2}{2} +..$~, the electric dipole moment per unit volume of the electron system can be cast in the form $P_e = P_e^{(0)}+\epsilon_0 \chi_e \mathcal{E}$, where $\chi_e =\chi_e^{(1)}+\chi_e^{(2)}\mathcal{E}+..$ is the nonlinear ac electric susceptibility of the electron system. By inserting this form into Eq.~(\ref{eq:deltaq}), a current in the capacitor plate can be written as 
\begin{eqnarray}
i(t)& =&\frac{d(C_0  u(t)  + \Delta q)}{dt} \label{eq:i_Vac_0} \nonumber \\
& =& C_0  \left(1+\eta \chi_e^{(1)}+ \eta  \chi_e^{(2)} \frac{2 u(t)}{D}+.. \right)\frac{du(t) }{dt} 
\label{eq:i_Vac}
\end{eqnarray}
with  $\eta =\Delta z_n/D$. The image current is caused by the electric susceptibility of the electron system induced by the MW-induced population of the excited states. The linear susceptibility is given by $\chi_e^{(1)}=-\frac{ en_s }{\varepsilon_0} \frac{d\rho_{nn}}{d \mathcal{E}}|_{u=0}  $ and can be rewritten as \begin{equation}
\chi_e^{(1)}=-\frac{\alpha en_s }{\varepsilon_0} \left( \frac{\partial\rho_{nn}}{\partial\omega_{1n}} \right),
\label{eq:chi}
\end{equation}
\noindent using an approximated linear dependence $\alpha=d\omega_{1n}/dE_\perp$. Eq.~(\ref{eq:i_Vac}) states that the relative change in capacitance $\Delta C /C_0 =\eta \chi_e^{(1)} $ due to the linear part of $\chi_e$ is equivalent to inserting a dielectric slab with thickness $\Delta z_n $ and susceptibility $\chi_e^{(1)} $ and the nonlinear part of $\chi_e$ produces higher harmonic time-dependence. Note that, unlike the current $i$ which is proportional to the total number of electrons $N_e=n_s S$ in the capacitor, the relative change $\Delta C/C_0$ is proportional to the electron density $n_s$. Thus, the required number of the electrons for detection scales down with the size of the capacitor C employed for the detection. The sensitivity of the measurement can be improved, for example, by including   the capacitor $C$ in an $LC$-circuit~\cite{Gonzalez-ZalbaNatCom2015}. As mentioned earlier, employment of a parallel-plate capacitor of a much smaller size than used here is not practical due to the capillary action of liquid helium~\cite{CollPRB2017}. It would be preferable to use, for example, a coplanar-plate capacitor covered by a thin helium film. For the detection of the quantum states of a single electron, the linear dimensions of such a capacitor can be reduced to $\mu$m-size. This would also give about three-order of magnitude enhancement in the value of $\eta$, which is inversely proportional to the capacitor gap $D$. Considering a capacitor with $S=10\times 10$~$\mu$m$^2$ and $D=1$~$\mu$m, and using experimentally determined $\alpha\sim 5$~MHz/(V/m), for a single electron we obtain $\Delta C/C_0\sim 30 (\partial\rho_{22}/\partial \omega_{1n})$, where $\omega_{1n}$ is in MHz. For a single electron localized in a dc electrostatic trap we can assume an intrinsic broadening of the transition line $\gamma\sim 1$~MHz, which gives us $\Delta C/C_0>> 1$. This crude estimate ignores experimental effects, such as stray capacitance, lateral motion of an electron in the trap, and inhomogenious broadening of the transition line. However, bearing in mind that employment of $LC$-circuits allows for determination of relative changes $\Delta C/C_0$ to part-per-million~\cite{Gonzalez-ZalbaNatCom2015}, the proposed scheme of detection is promising.    

Once the detection of the excitation of the Rydberg state is realized for a single electron, we could transform it to a non-destructive readout of the spin state of a single electron with the help of a magnetic field gradient. A current running through a superconducting wire in the vicinity of the trapped electrons~\cite{SchuPRL2010} or a ferromagnet under an external magnetic field~\cite{TokuraPRL2006} can create a local magnetic field gradient. Thanks to the field gradient, the electron feels a different magnetic field depending on the Rydberg state, which allows spin-selective excitation of the Rydberg state. The detailed schematics for the detection of the spin state of a single electron and its manipulation are beyond the scope of this paper and will be discussed elsewhere.

In summary, we have proposed a new method, the image-charge detection, for spectroscopic study of the Rydberg states of surface electrons on liquid helium. The method is demonstrated by measuring the image current induced in the capacitor circuit by a pulse-modulated MW excitation. The method is simple, does not require expensive devices such as a cryogenic hot-electron (InSb) bolometer, and can be readily employed using a conventional lock-in amplifier. Moreover, it is demonstrated that the image-charge detection provides other advantages over the conventional methods, such as the ability to do spectroscopic studies of the high-lying Rydberg states. In addition to SE, this method can be potentially useful to study radiation-induced intersubband transitions in 2D electron gas in semiconductor heterostructures, where energy structure of electrons trapped at the interface of two solid materials has many similarities with that of SE on liquid helium~\cite{AndoRevModPhys1982}. We also discussed the possibility of using this method to detect the spin state of a single electron, which opens a new pathway for using spin states of SE  on liquid helium for quantum computing.

This work was supported by JST-PRESTO (Grant No. JPMJPR1762) and an internal grant from Okinawa Institute of Science and Technology (OIST) Graduate University. 
We are grateful to V.~P. Dvornichenko for providing technical support.

\end{document}